\documentclass[conference]{IEEEtran}
\IEEEoverridecommandlockouts
\usepackage{xurl}
\usepackage{hyperref}
\usepackage{siunitx,booktabs}
\usepackage{cite}
\usepackage{amsmath,amssymb,amsfonts}
\usepackage{algorithmic}
\usepackage{graphicx}
\usepackage{textcomp}
\usepackage{xcolor}
\usepackage[linesnumbered,ruled,vlined]{algorithm2e}
\def\BibTeX{{\rm B\kern-.05em{\sc i\kern-.025em b}\kern-.08em
    T\kern-.1667em\lower.7ex\hbox{E}\kern-.125emX}}
\begin{document}

\title{LNMesh: Who Said You need Internet to send Bitcoin? Offline Lightning Network Payments using Community Wireless Mesh Networks}


\author{\IEEEauthorblockN{
Ahmet Kurt,
Abdulhadi Sahin,
Ricardo Harrilal-Parchment, and
Kemal Akkaya}
\IEEEauthorblockA{Dept. of Electrical and Computer Engineering, Florida International University, Miami, Florida 33174\\
Email: \{akurt005, asahi004, rharr119, kakkaya\}@fiu.edu}
}

\maketitle

\begin{abstract}
Bitcoin is undoubtedly a great alternative to today's existing digital payment systems. 
Even though Bitcoin's scalability has been debated for a long time, we see that it is no longer a concern thanks to its layer-2 solution Lightning Network (LN). LN has been growing non-stop since its creation and enabled fast, cheap, anonymous, censorship-resistant Bitcoin transactions. However, as known, LN nodes need an active Internet connection to operate securely which may not be always possible. For example, in the aftermath of natural disasters or power outages, users may not have Internet access for a while. Thus, in this paper, we propose LNMesh which enables offline LN payments on top of wireless mesh networks. Users of a neighborhood or a community can establish a wireless mesh network to use it as an infrastructure to enable offline LN payments when they do not have any Internet connection. As such, we first present proof-of-concept implementations where we successfully perform offline LN payments utilizing Bluetooth Low Energy and WiFi. For larger networks with more users where users can also move around, channel assignments in the network need to be made strategically and thus, we propose 1) minimum connected dominating set; and 2) uniform spanning tree based channel assignment approaches. Finally, to test these approaches, we implemented a simulator in Python along with the support of BonnMotion mobility tool. We then extensively tested the performance metrics of large-scale realistic offline LN payments on mobile wireless mesh networks. Our simulation results show that, success rates up to \%95 are achievable with the proposed channel assignment approaches when channels have enough liquidity. 
\end{abstract}


\section{Introduction}

When Bitcoin \cite{nakamoto2008bitcoin} was introduced in 2009, it created a big sensation in the world as it was first of its kind. Since then, a lot of different cryptocurrencies were proposed. Today, cryptocurrencies can be used to pay for goods and services similar to using cash or credit cards. However, none of them could replace or supersede Bitcoin in usage or market capitalization. Current market conditions still implie that it will stay the same way \cite{saiedi2021global}. 

However, Bitcoin suffers from very low transaction per second (TPS) which limits its usability on large scale. There have been numerous proposals to increase its scalability such as \textit{block size increase}, \textit{Schnorr signatures}, \textit{side chains} and \textit{layer-2 networks} \cite{zhou2020solutions}. Among all, layer-2 networks is by far the most promising solution as shown with the success of the Lightning Network (LN) \cite{poon2016bitcoin} which grew exponentially over the years reaching 16,000 public nodes worldwide. 

LN was implemented in 2017 with the aim of decreasing the load on the Bitcoin blockchain by facilitating the transactions on its decentralized network which enables almost free and instant Bitcoin payments. It works by processing the payments \textit{off-chain} meaning payments are not recorded on the Bitcoin blockchain. In order to transact on LN, users need to open at least one LN channel to one of the nodes in the network in advance and put some funds in the channel. With the help of these technologies, now we have Bitcoin even on IoT and mobile devices \cite{kurt2021enabling}. 

These developments along with the decentralized nature of Bitcoin offer opportunities to explore whether digital payments can be realized when there is no Internet connection. This is particularly crucial for the cases when there are natural disasters such as hurricanes or earthquakes causing power and Internet outages while people in communities still need to interact and make payments. Indeed, this was a particular issue after Hurricane Irma in 2017 when people in South Florida did not have Internet for weeks\footnote{\url{https://mashable.com/article/hurricane-irma-power-outage-florida}} while they still needed to make payments for gas, groceries, repairs and other basic needs. In such cases, a method to enable LN payments without needing an active Internet connection (i.e., offline LN payments) among people as well as business owners will greatly benefit the community.

Thus, in this paper, we propose using LN on top of community wireless mesh networks to enable sending/receiving offline LN payments between the members of the mesh network without needing any Internet connection. In this way, until users get back online, they can transact using their existing LN channels. LN protocol allows such offline payments to settle since the payments are off-chain and not recorded to the Bitcoin blockchain. Thus, as long as nodes can communicate with each other through wireless technologies such as WiFi or Bluetooth, they can perform offline LN payments.

To realize offline LN payments, we first conducted a feasibility study where we implemented a proof of concept using 8 Raspberry Pi devices. The setup involves 1) creating a wireless mesh network among the Pis, 2) forming an LN topology on top of the mesh, 3) cutting nodes' Internet connection, and 4) performing offline LN payments. For wireless technologies, we chose Bluetooth Low Energy (BLE) and WiFi. Initially, we created an \textit{IP-over-BLE Star Network} to test the feasibility of the concept. For the WiFi, we created a full mesh topology by placing the Pis around our university campus building. On top of this mesh topology, we created 3 different LN topologies to test the performance of offline LN payments. Our conclusion from these experiments is that, having an established wireless mesh network, offline LN payments are indeed possible with reasonable success rates and delays.

However, scaling this concept to a large-scale community mesh network creates many challenges to tackle, such as opening the LN channels among the members of the mesh network. For example, it is not realistic to expect every user of the mesh network to open channels to every other user in the mesh. Even though that would result in a very well connected LN topology, 1) it would require a lot of monetary investment from the users, and 2) some channels would be redundant as users can utilize multi-hop links without needing direct channels to every other user in the network. Additionally, users of the mesh network are expected to be moving around (i.e., mobile). Thus, channels need to be opened strategically and this boils down to the \textit{LN channel assignment problem in mobile wireless mesh networks with minimized investment costs}. To the best of our knowledge, this problem is not studied in the literature yet. 

Before deciding how to open the channels, we first capture the mobility patterns of the users in the mesh network. Basically, we observe the movements of the users in a given area for a while, and record their positions at different times essentially getting their mobility traces. If any two nodes seem to be around each other for a significant amount of time, we include them in the \textit{mobility-aware mesh topology}. This new topology represents a probabilistic virtual topology taking into account the users' mobility patterns. Here, we propose two different approaches to assign the channels. Both of the approaches are focused on reducing the number of links in the \textit{mobility-aware mesh topology} and forming a spanning tree. Our first approach is based on connected dominating set (CDS) concept. CDS creates dominator and dominatee nodes where dominators form the network backbone. Knowing which nodes are inside and outside of the network backbone, we can assign one LN channel from each dominatee to one of the dominators and one channel among the dominators themselves. Second approach calculates a uniform spanning tree instead of a CDS to assign one channel for each edge of the tree. 

To assess the effectiveness of our channel assignment approaches, we implemented a simulator in Python and performed various experiments. We utilized real mobility scenarios created with the BonnMotion tool \cite{aschenbruck2010bonnmotion} using a real map of our university campus building and its surroundings. Using the simulator, we can change the total investment amount in the network, number of users in the network, number of transactions at each time frame, wireless coverage of the users and more. Our results show that with enough liquidity on the channels and using our channel assignment approaches, we can reach up to \%95 success rate on all payments performed during a day. We also report the success rates for more challenging cases with a lot users and very low total investment.

The rest of the paper is organized as follows. In Section \ref{sec:relatedwork}, we provide the related work. Section \ref{sec:feasibilitystudy} describes our proof of concept implementation for realizing offline LN payments in wireless mesh network. The proposed channel assignment approaches are explained in Section \ref{sec:approaches}. In Section \ref{sec:simulations}, we present detailed performance analysis of the proposed channel assignment approaches using the simulator we implemented. Finally, we conclude the paper in Section \ref{sec:conclusion}.

\section{Related Work}
\label{sec:relatedwork}

While there has been a lot of work on the integration of Bitcoin with IoT devices \cite{kurt2021lngate, fernandez2018review}, the concept of fully offline Bitcoin payments received little interest. The closest work to ours is from Myers \cite{lot49}. The author proposes a protocol called Lot49 that aims to incentivize message senders in a mesh network using LN-like payment channels. While the work looks interesting in the direction to possibly enable offline LN payments, it has too many requirements and assumptions that are unlikely to be feasible. For example, the protocol requires Schnorr signatures and \texttt{SIGHASH\_NOINPUT} flag to be adopted by the Bitcoin community, changes to the Bitcoin scripts of LN, and many different types of nodes to be setup in the mesh. Additionally, there is no proof of concept implementation of the protocol. In contrast, our work does not need any modifications to the LN or Bitcoin protocols, and works without setting up so many different types of nodes. 

It is also worth mentioning that, the idea of using mesh networks to enable offline Bitcoin and LN payments were mentioned at several news articles\footnote{\url{https://bitcoinmagazine.com/technical/making-bitcoin-unstoppable-part-one-mesh-nets}}\footnote{\url{https://bitcoinmagazine.com/technical/from-isp-to-p2p-how-mesh-networks-take-bitcoin-off-the-grid}}. However, they do not offer any actual solution. There are also commercialized mesh networking solutions such as Locha Mesh\footnote{\url{https://locha.io/}} that might be used for sending Bitcoin offline. However, users need to purchase special nodes and the system is geared more towards communication or chat. Additionally, the open source development of the software seems to be dormant as of now. 

A practical problem in this context is to temporarily accommodate offline LN nodes. While there was no scientific study, there have been discussions in the Bitcoin and Lightning community about how offline LN nodes can receive payments. Such payments are called asynchronous (async) payments. A thread on the Lightning-dev mailing list was started by Matt Corallo to discuss the possible solutions\footnote{\url{https://lists.linuxfoundation.org/pipermail/lightning-dev/2021-October/003307.html}}. According to him, Point Time Locked Contracts\footnote{\url{https://bitcoinops.org/en/topics/ptlc/}} (PTLCs) which were proposed to replace HTLCs in LN, could be the best option. PTLCs use public key for locking and a corresponding signature for unlocking in contrast to HTLCs' hash and preimage combination. Since PTLCs are not yet implemented in LN; a partial solution, trampoline relays were proposed\footnote{\url{https://github.com/ACINQ/eclair/pull/2435}}. These relay nodes are managed by third parties and can temporarily hold the payments until the offline recipient node comes back online. However, this approach does not address payment sending between two offline LN nodes. In our approach, we enable offline-to-offline payments utilizing wireless mesh networks.

There were some efforts on enabling offline but on-chain Bitcoin transactions as well which works based on the concept of online coin preloading. The first feasible idea was proposed by Dmitrienko et al. \cite{dmitrienko2017secure} where authors proposed utilizing offline wallets leveraging secure hardware. Later, Takahashi et al. \cite{takahashi2020short} slightly improved on this idea by removing the external trusted time-stamp server in the design. However, both approaches need wallets produced by trustworthy manufacturers which raises security concerns. Additionally, the approach requires changes to the Bitcoin protocol.

Researchers also explored other blockchains for realizing offline cryptocurrency payments. For example, Rawat et al. \cite{rawat2022offline} explored whether IOTA blockchain can be used to perform offline payments. They concluded that current IOTA blockchain cannot accommodate the desired offline payments without significant modifications to its protocol. A more concrete solution called \textit{DelegaCoin} was proposed by Li et al. \cite{li2021offline} whose main idea is to utilize Trusted Execution Environments (TEEs) for secure offline delegation of coins without interacting with the blockchain. However, this approach requires additional entities to be set up (i.e., TEEs) which may not be practical for all users. In contrast to these works, we propose using LN on top of existing wireless mesh networks without needing to modify existing blockchain protocols or setting up new entities.

\section{Feasibility Study}
\label{sec:feasibilitystudy}

To demonstrate the feasibility of the approach, we first created a proof of concept implementation. This section provides the details of this implementation and the corresponding results. The proof of concept will shed light on how to design such a system in large-scale which will be tackled in the next sections.

\subsection{Implementation Environment}
Our implementation utilizes 8 Raspberry Pi 4 Model B each having a 64GB SD card loaded with 64-bit Raspberry Pi OS (Raspbian). Each of these Raspberry Pis support Bluetooth Low Energy (BLE) and WiFi interfaces with dual-band IEEE 802.11b/g/n/ac and Bluetooth 5.0. For the Bitcoin nodes, we installed \textit{bitcoind v0.23.0} \cite{bitcoind} and for the LN nodes, we used \textit{Core Lightning v0.11.2} \cite{clightning}. All the transactions were performed on Bitcoin's Testnet. To fund the LN nodes with Testnet Bitcoins, we used a faucet\footnote{\url{https://coinfaucet.eu/en/btc-testnet/}}. Note that in LN, channels need to be opened in advance and transactions can be sent to recipients over multi-hop routes. If there is no route to the recipient, the payment will not go through.

\subsection{Bluetooth Implementation and Results}

We first tested the offline LN payments on a BLE setup to have an initial proof of concept. The setup consists of 8 Pis where one of Pis is the master and the rest are slaves. LN works with TCP/IP thus BLE alone cannot be used to perform LN operations between the nodes. BLE setup needs to be adjusted such that the TCP/IP can work over BLE. To do this, we did some configuration on the Pis such as installing \textit{bluez-tools}, creating a personal area network, configuring the master node and slave nodes and more. However, doing only that is not enough as we are also trying to create a mesh network. Such a setup is then called IP mesh over BLE with no known implementation to-date. As we are only trying to test the feasibility of the concept, we instead created a \textit{IP-over-BLE Star Network}. Basically, one of the Pis were designated as the master and other 7 connected to it as slaves. The full details and step-by-step guide of this implementation is given at our GitHub page at \texttt{\url{https://github.com/startimeahmet/LNMesh/tree/main/BLE_star}}. 

\begin{figure}[h]
    \centering
    \vspace{-4mm}
    \includegraphics[width=0.8\linewidth]{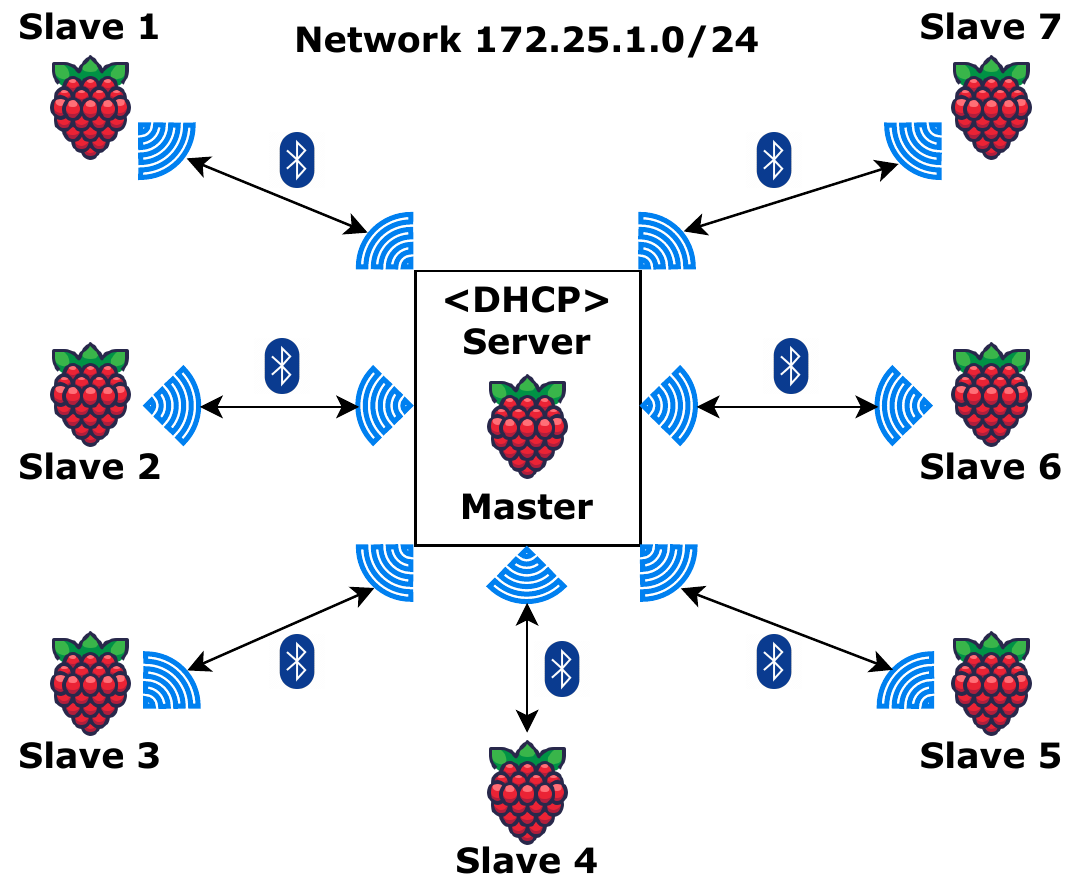}
    \vspace{-4mm}
    \caption{Topology of the IP-over-BLE star network.}
    \vspace{-2mm}
    \label{fig:BLE_setup}
\end{figure}

Once this IP-over-BLE star network setup was ready, the next step was to open the LN channels. For this, we again chose a star topology where all slaves opened a channel to the master. The physical placement of the slaves around the master was arbitrary. The virtual topology of this setup is illustrated in Fig. \ref{fig:BLE_setup}.

Once the Bitcoin and LN nodes of all Pis were synced, we cut the Internet connection on all the Pis at the same time to take them all offline. This is a critical part in the experiments as our motivation is to test offline LN payments. In our tests, we realized that, cutting the Internet connection puts the Bitcoin and LN nodes into \textit{searching for peers} mode. Both nodes still keep running but lose their existing peers because of having no Internet connection. Thus, we just went ahead and connected each LN node on the slave Pis to the LN node on the master Pi using their local IP addresses \texttt{172.25.1.xxx} with the command \texttt{lightning-cli connect pubkey@IP:port}. In a real deployment, this process can be automated with a script which will automatically get the local IPs of the peers and force a reconnect. This way, we did not need to make any modifications on the LN or Bitcoin software to make offline LN payments work.

For the experiments, we executed 100 payments in total where we randomly selected two nodes among the 8 nodes to send payments to each other. This was automated with a bash script we wrote. Since the LN channels were opened to form a star topology, payments can either be between the slaves and the master (direct payments) or between the slaves (1-hop payments). All executed payments were successful. The average delay was around 1.5 seconds for direct payments and around 2.5 seconds for 1-hop payments.

\subsection{WiFi Implementation and Results}

Next, we tested the concept with WiFi as it will offer longer coverage. Unlike the BLE case, we created a full mesh network (i.e., not through a master) for the WiFi experiments. Thus, it involves a more complicated setup since the nodes have to be placed at a distance from each other to create a mesh topology. To create the WiFi mesh network, we used \textit{batman-adv} \cite{batmanadv} which is a routing protocol specifically designed for mobile ad-hoc networks and is part of the Linux kernel. One of the Pis were setup as the mesh gateway to provide Internet connectivity to the rest of the Pis in the mesh. For that, we used an additional generic Wireless USB adapter on the gateway to connect it to the Internet when necessary. This Internet connectivity was only used to sync the Bitcoin and LN nodes on the Pis. The mesh network was created inside our university building and the nodes do not necessarily have line of sight propagation between each other. The full details and step-by-step guide of this implementation is given at our GitHub page at \texttt{\url{https://github.com/startimeahmet/LNMesh/tree/main/WIFI_mesh}}.

\begin{figure}[ht]
    \centering
    \vspace{-1mm}
    \includegraphics[width=0.6\linewidth]{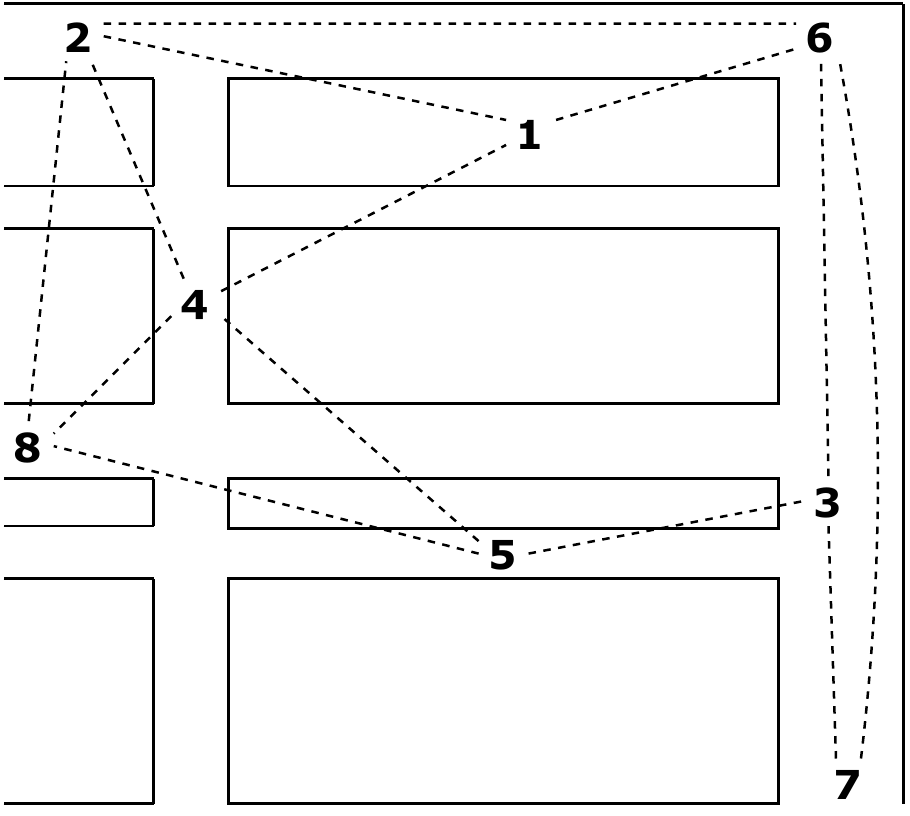}
    \vspace{-3mm}
    \caption{Placement of the Pis in the building and the resulting mesh topology for the WiFi mesh setup. Dashed lines show the mesh links.}
    \vspace{-2mm}
    \label{fig:mesh_topology}
\end{figure}

In Fig. \ref{fig:mesh_topology}, we show the topology of the nodes inside FIU Engineering Center building and their connections to each other on the mesh level. The rectangles in the figure represent the rooms and walls. The dashed lines on the figure represent the neighbors of each node on the mesh network which were identified with the command \texttt{sudo batctl n} where \textit{batctl} is the configuration and debugging tool for \textit{batman-adv}.

\begin{figure*}[htb!]
    \centering
    \includegraphics[width=0.95\linewidth]{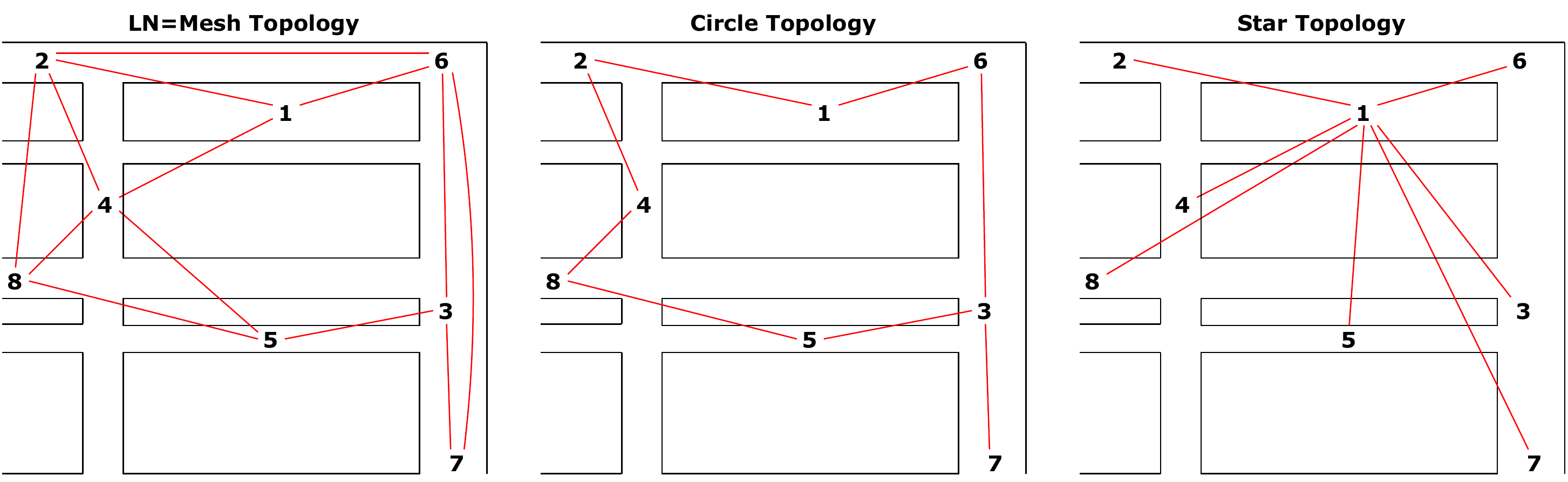}
    \vspace{-3mm}
    \caption{Illustrations of different LN topologies tested on the WiFi mesh setup. Red lines show the LN channels.}
    \vspace{-5mm}
    \label{fig:LN_topologies}
\end{figure*}

Next, we created several LN topologies by opening channels between the nodes. These topologies can be seen in Fig. \ref{fig:LN_topologies}. First topology is the same as the mesh topology where each node has a channel with its neighboring nodes. Second one is a circular topology where the channels form a complete loop. Finally, the third one is a star topology where all nodes have a channel to the gateway node. Same as the BLE case, we cut the Internet connection of the nodes after the channels were opened and peered them with each other on LN using their local IPs \texttt{192.168.199.xxx}.

For all these topologies, again similar to the BLE case, we performed 100 payments between two randomly selected nodes among the 8 nodes we had. Payment amounts were set to 5,000 satoshi and all channel capacities were initially set to 100,000 satoshi. The procedure of choosing the nodes randomly and executing the payments between the nodes was automated using a bash script. Randomization was done using fixed seed values to prevent different nodes to be selected for different experiment runs. The results of these experiments are presented in Table \ref{tab:WiFi_results}.

\begin{table}[h]
 \vspace{-3mm}
  \begin{center}
  \vspace{-2mm}
    \caption{WiFi Mesh Experiment Results}
    \vspace{-2mm}
    \label{tab:WiFi_results}
    \resizebox{0.85\linewidth}{!}{
    \begin{tabular}{|c|c|c|c|}
    \hline
    \textbf{Topology} & \textbf{Success} & \textbf{Average} & \textbf{Total Number}  \\
    \textbf{}         & \textbf{Rate}             & \textbf{Delay} & \textbf{of Channels}  \\ \hline
    \textbf{LN=Mesh} &  \%95   &  5.95 sec  &  13 \\ \hline
    \textbf{Circle} &  \%53  & 6.31 sec &  8 \\ \hline
    \textbf{Star} &  \%60  & 4.76 sec  &  7 \\ \hline
    \end{tabular}
    }
   \vspace{-4mm}
  \end{center}
\end{table}

Highest success rate was observed with the LN=Mesh topology which makes sense as it has more channels than the other two thus more routing options/less failures. Circle topology on the other hand has the lowest success rate and the most delay which again makes sense because routing options are limited and payments have to go through more hops. Finally, we see that the star topology performs slightly better than the circle with the least delay among all options which is logical as the payments are either direct or 1-hop payments. In general, these results show us that while offline payments are possible, the way the channels are assigned makes a major impact on the success rate when considering mesh topologies. Therefore, we tackle this problem next.

\section{Channel Assignment Approaches}
\label{sec:approaches}

In the previous section, we showed that offline LN payments can indeed be realized in a wireless mesh network utilizing WiFi or BLE. Now in this section, we will present techniques to extend such a setup to a large number of users and channels, which can be used in a real-life application within a community when they all lose Internet connection at the same time due to a disaster such as a hurricane/earthquake etc. In this direction, we first describe how to handle large-scale mesh topologies that change over time (i.e., mobile) and then offer two different channel assignment techniques for mobile community wireless mesh networks.

\subsection{Problem Motivation and Handling Mobility}
\label{sec:handling_mobility}

Opening and closing LN channels require Internet connection since they are on-chain Bitcoin transactions broadcast to the Bitcoin network. This means that the users have to open their channels before they go offline. Therefore, our problem boils down to the following: Given a community of people moving around in a neighborhood during a day, how to decide who opens a channel to who so that the overall success rate of the payments in the network will be satisfactory when people do not have access to the Internet?

A good starting point for deciding how to open channels in a wireless mesh network is by looking at the individual mesh connections of the users. We could open an LN channel for every mesh link and have an LN topology that is the same as the mesh topology. However, this will not be possible since we are dealing with a mobile topology that changes over time. Additionally, opening a channel for every single mesh link would be very costly as it would cause too many channels to be opened that would require higher initial investment from all the users. Instead, a more optimized way is to look at users' mobility patterns and make channel assignments based on these patterns to reduce the number of channels opened. Mobility pattern of a user can be interpreted as how frequently the user move around other users and how often it gets close to others. In real life, this corresponds to how often people interact with each other or with stores, gas stations and markets around them.

\vspace{-2mm}
\SetAlCapNameFnt{\small}
\SetAlCapFnt{\small}
\begin{algorithm}[h]
\small
\caption{Mobility-Aware Mesh Topology Generation}
\label{algo:algo1}
\newcommand{\hrulealg}[0]{\vspace{1mm} \hrule \vspace{1mm}}
\newcommand\mycommfont[1]{\footnotesize\ttfamily\textcolor{black}{#1}}
\SetCommentSty{mycommfont}
\SetAlgoLined

\SetKwInOut{KwIn}{Input}
\SetKwInOut{KwOut}{Output}

\KwIn{Mesh distances $data$ with following 4 columns: [$source$, $target$, $time$, $distance$]}
\KwOut{Mobility-aware mesh topology $G_{mesh}$}

\hrulealg

\SetKw{KwIn}{in}

\textbf{define} $k$\tcc*{the metric to include the nodes in the new topology}
\textbf{define} $d$\tcp*{wireless coverage of the nodes}
\textbf{initialize} empty $G_{mesh}$\;
\For {each ($source$, $target$) \KwIn $data$}{
    $count$ = 0\tcc*{how many times two users get close enough}
    \For {each $time$ \KwIn $data$}{
        \lIf {$distance \leq d$}{$count$ += 1}
    }
    \If {$count \geq k$}{
        $G_{mesh}$.add\_edge($source$, $target$)\;
    }
}
\textbf{return} $G_{mesh}$\;
\end{algorithm}
\vspace{-4mm}

To model these mobility patterns, we need to analyze the mobility behavior of the users for a period of time within a specific neighborhood. For instance, if we can collect mobility traces for certain users during a day in a neighborhood, these traces can then be merged to create instant mesh topologies at certain times. In other words, we can get a snapshot of mesh topologies at certain intervals and then identify the most probable neighbors of each user in average. This information can be combined to create an average wireless mesh topology of users during a typical day. This topology will give us a hint on how to assign channels. We came up with Algorithm \ref{algo:algo1} to create this topology.

Algorithm \ref{algo:algo1} is pretty straightforward in the sense that it gets as input the distance information between all possible users in the mesh at different times of the day and outputs the \textit{mobility-aware mesh topology}. For all given times (line 6), it checks the number of occurrences any two users \textit{(source, target)} (line 4) get close to each other less than the distance $d$ (line 7). Here, $d$ stands for the wireless transmission range of the users. If the number of occurrences is at least $k$ times (line 9), we decide that these two users frequently interact with each other and add them to the mobility-aware mesh topology (line 10).

\subsection{Problem Definition and Formulation}
Now that we have the mobility-aware mesh topology which is fixed (i.e., does not change with time anymore), the next problem is how to assign the LN channels on it. The problem of channel assignment for mobile ad-hoc networks can be modeled based on one of the variants of the well-known transportation problem \cite{appa1973transportation}. Mainly, the goal in this problem is to move certain supplies from suppliers to warehouses through certain routes that come with specific costs. In our case, the supplies would correspond to payments, suppliers and warehouses would be users and the cost of the route would be associated with the capacity (i.e., higher capacity channels are cheaper while lower capacity ones are expensive). However, in our case, there is an additional problem since we are trying to prune the resultant topology to reduce the number of channels opened. This problem is related to the connectivity of the topology. From a given topology which represents an undirected graph $G$ $(V, E)$ where $V$ is the set of users and $E$ are the wireless links among them, we need to guarantee that the selected links eventually form a connected topology so that any payment can reach any destination. This connectivity constraint makes the problem intractable for sensor networks as shown in \cite{sir2011optimization}. Therefore, we opt to follow approaches which can offer connected topologies with minimal number of links. Specifically, we will prune the topologies created by Algorithm \ref{algo:algo1}.

To this end, one easy solution is to utilize \textit{spanning trees} from graph theory which has been widely used in the context of mobile ad-hoc networks for various purposes \cite{kurt2021distributed}, \cite{baala2003self} since they guarantee connectivity with minimum number of links. In graph theory, spanning tree of a given undirected graph $G$ is a subgraph of $G$ that includes all the vertices of $G$ with minimum number of edges. That means, if $G$ has $n$ vertices, a spanning tree of $G$ has $n-1$ edges.

Finding the minimum spanning tree (MST) has been well studied, and there are polynomial time algorithms such as Kruskal's or Prim's \cite{wu2004spanning} to find the MST if each link has an associated cost (i.e., weight). In our case, the different link costs used for an MST are not applicable since once created, each link is equal in terms of serving as an LN channel. In other words, each link will have the same cost. Therefore, any spanning tree would be applicable to our solution. Nevertheless, we recognize that even if the number of links (and thus the number of LN channels to be opened) will be same for any spanning tree topology, the degree of nodes (i.e., the number of neighbors for a node) may differ significantly. In other words, the variance of the node degrees may change from topology to topology. 

Therefore, we plan to pursue two approaches to form our spanning trees: Using minimum connected dominating set (MCDS) \cite{guha1998approximation} and uniform spanning tree (UST) \cite{propp1998get} concepts. Next two sections are dedicated to these approaches.

\vspace{-2mm}
\begin{algorithm}[h]
\small
\caption{MCDS-based LN Topology Generation}
\label{algo:algo2}
\newcommand{\hrulealg}[0]{\vspace{1mm} \hrule \vspace{1mm}}
\newcommand\mycommfont[1]{\footnotesize\ttfamily\textcolor{black}{#1}}
\SetCommentSty{mycommfont}
\SetAlgoLined

\SetKwInOut{KwIn}{Input}
\SetKwInOut{KwOut}{Output}

\KwIn{Mobility-aware mesh topology $G_{mesh}$}
\KwOut{MCDS-based LN topology $G_{LN_{MCDS}}$}

\hrulealg

\SetKw{KwIn}{in}

$G_{MCDS}$ = compute\_MCDS($G_{mesh}$)\tcc*{compute the MCDS graph using any known MCDS algorithm}
$G_{MCDS_{nocycles}}$ = compute\_MST($G_{MCDS}$)\tcc*{compute MST of MCDS graph to remove possible cycles}
$G_{LN_{MCDS}}$ = $G_{MCDS_{nocycles}}$\tcc*{initialize LN topology}
\For {each $node$ \KwIn $G_{mesh}$}{
    \If {$node$ not \KwIn $G_{MCDS}$}{
        $neighbors$ = $G_{mesh}$.neighbors($node$)\;
        $possibilities$ = $neighbors$ $\cap$ $G_{MCDS}$\;
        $G_{LN_{MCDS}}$.add\_edge($node$, $possibilities[0]$)\tcc*{choose the first possibility}
    }
}
\textbf{return} $G_{LN_{MCDS}}$\;
\end{algorithm}
\vspace{-5mm}

\subsection{Minimum Connected Dominating Set Approach}

As mentioned before, we are trying to assign minimum number of channels in the network to reduce the burden on the users to open many channels which requires monetary investment. In this direction, we can first find the core vertices of the mobility-aware mesh topology graph and then follow this core to assign channels. The rationale behind this is to form a topology such that any node can reach any other node through this core with relatively shorter paths. This can indeed be measured through metrics such as \textit{closeness centrality} and \textit{betweenness centrality} \cite{GOLBECK201325} from graph theory. Shorter paths increase the success rates because there are less channels on the payment path that might cause the payment to fail. The other motivation behind determining a core is to allow any potential businesses to be part of the core if they offer a wireless node to interact with the users. 

Core vertices in a topology can be found with a minimum connected dominating set (MCDS) algorithm. Connected dominating set $D$ (i.e., dominator nodes) of a graph $G$ has the following properties:
\begin{itemize}
    \item $D$ forms a connected subgraph of $G$.
    \item If a vertex in $G$ is not in $D$ (i.e., a dominatee node), it is adjacent to a vertex in $D$. 
\end{itemize}
Thus, MCDS finds a $D$ with minimum number of vertices. We present the pseudocode of our MCDS-based LN topology generation algorithm in Algorithm \ref{algo:algo2}.

\begin{figure}[h]
    \centering
    \vspace{-2mm}
    \includegraphics[width=\linewidth]{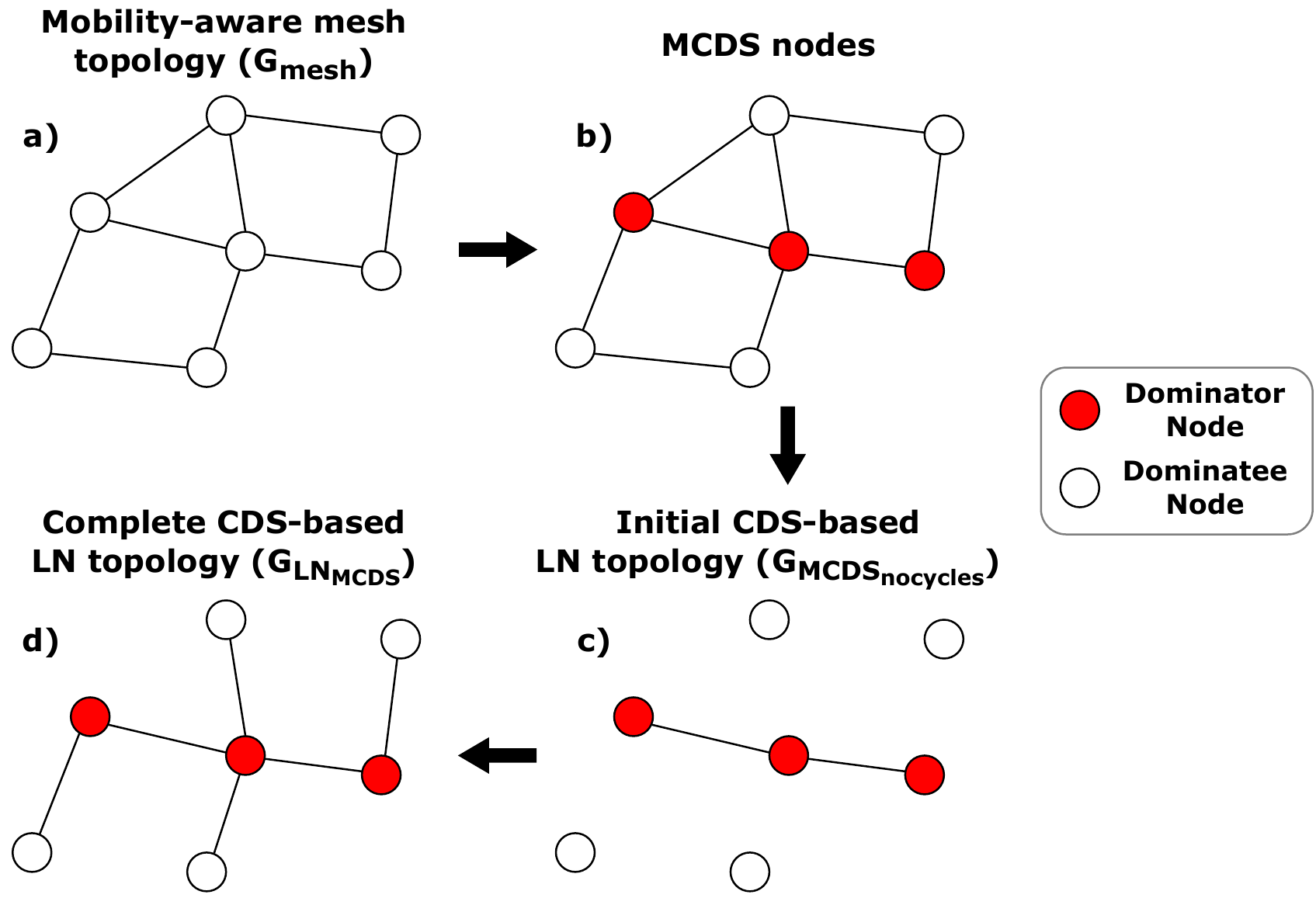}
    \vspace{-7mm}
    \caption{An illustration of the workflow of Algorithm \ref{algo:algo2}}
    \vspace{-2mm}
    \label{fig:MCDS}
\end{figure}

Algorithm \ref{algo:algo2} takes the mobility-aware mesh topology graph as an input and returns a processed version with fewer number of links. We first compute the MCDS graph of the mobility-aware mesh topology graph (line 1). For finding the MCDS, any known algorithm in the literature can be used \cite{guha1998approximation}. However, when we find it, we remove any cycles within this MCDS by computing its MST (line 2). This resulting graph is also the starting point to create our LN topology thus we initialize the LN topology graph to the MCDS graph with no cycles (line 3). The idea is to enlarge the MCDS graph with no cycles until each dominatee node has a channel to one dominator node. To do this, we iterate over each node in the mobility-aware mesh topology graph that is not a member of the MCDS (i.e., dominatee nodes) and get their neighbor set (line 4-6). In this neighbor set, we check for the nodes that are in the MCDS (line 7). These nodes are potential candidates to open a channel to because they are already in the MCDS and have a link to the dominatee node in the mobility-aware mesh topology. We take the first candidate and add this (dominatee, dominator) pair to the LN topology graph (line 8). In this way, every node will be connected and there will not be any cycles, essentially another spanning tree. Once this process is done for all the dominatees, we output the LN topology graph. The way this algorithm works is also illustrated in Fig. \ref{fig:MCDS}.

\subsection{Uniform Spanning Tree Approach}

\begin{figure}[h]
    \centering
    \vspace{-3mm}
    \includegraphics[width=0.85\linewidth]{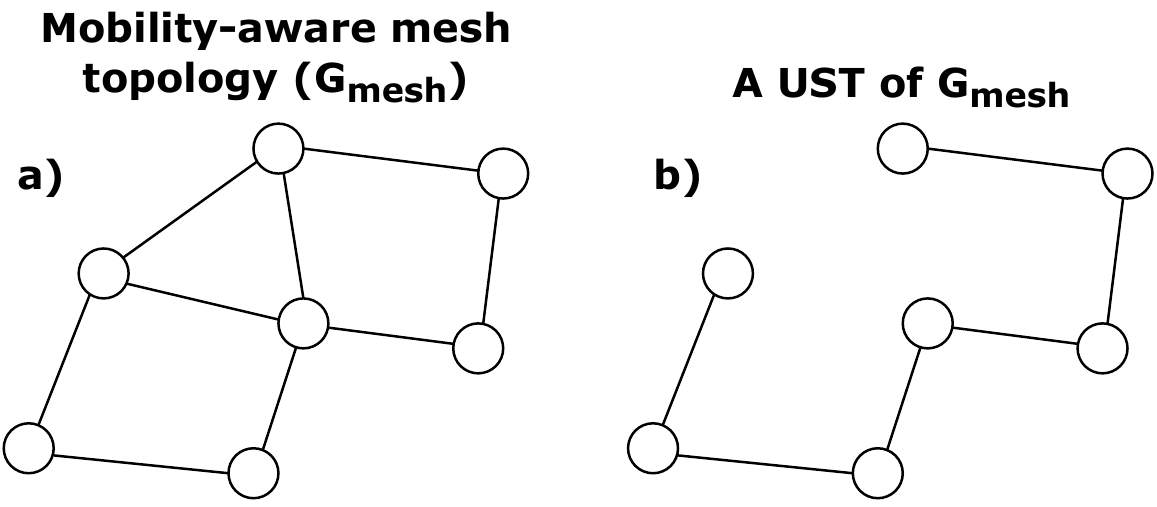}
    \vspace{-3mm}
    \caption{An example UST generated from $G_{mesh}$}
    \vspace{-4mm}
    \label{fig:MST}
\end{figure}

As an alternative to finding the MCDS, we can calculate spanning trees from the mobility-aware mesh topology graph using another method. As mentioned, since a graph can have multiple spanning trees, we can randomly select a spanning tree among all possible spanning trees of $G$ with equal probability. This is referred to as Uniform Spanning Tree (UST). USTs can be found using Wilson's algorithm which uses loop-erased random walks to generate them \cite{propp1998get}. We show an example UST in Fig. \ref{fig:MST} of the same graph in \ref{fig:MCDS}. The rationale behind this choice is to be able to compare random selection to a deliberate one (i.e., MCDS-based) and investigate the impact through experimentation.

\section{Simulations}
\label{sec:simulations}

To test our proposed approaches explained in Section \ref{sec:approaches}, we implemented a simulator in Python. It mostly utilizes the \textit{networkx} and \textit{pandas} libraries in Python to create the graphs of the mesh and LN topologies and perform operations on them. The full source code of our simulator is available in our GitHub page at \texttt{\url{https://github.com/startimeahmet/LNMesh/tree/main/simulator}}. 

\subsection{Implementing Mobility}
Before starting the simulations, we need to generate realistic mobility data to be able to create our mesh network. For this, we used the \textit{BonnMotion} software which is a mobility scenario generation and analysis tool written in Java \cite{aschenbruck2010bonnmotion}. Using this software, we can generate mobility scenarios based on popular mobility models within a given community. An example scenario generated by BonnMotion is shown in Table \ref{tab:scenario}. As can be seen, a scenario file created by BonnMotion has 4 columns in it which are [\textit{node, time, x, y}]. For example, row 694 in Table \ref{tab:scenario} is interpreted as follows: Node 1 was at coordinates (260.0223, 453.6815) at time 79.8913.

\begin{table}
\centering
\sisetup{group-digits=false,table-number-alignment=right}
\caption{An example scenario created by BonnMotion for the duration of 21,600 seconds (6 hours) for 100 nodes}
\vspace{-2mm}
\label{tab:scenario}
\begin{tabular}{@{} l *{4}{S[table-format=-1.5]} @{}} 
\toprule
     & {\textbf{node}} & {\textbf{time}} & {\textbf{x}} & {\textbf{y}} \\ 
\midrule
\textbf{row 1}   &  {0}  & {0}  &  640.9129 &  574.3036 \\ 
\textbf{row 2}   &  {0}  & 18.1457 &  618.9429 &  574.0308 \\
\textbf{row 3}   &  {0}  & 21.9939 &  618.9429 &  568.2012 \\
\textbf{.}  &  \textbf{.}  & \textbf{.} & \textbf{.} & \textbf{.} \\
\textbf{.}  &  \textbf{.}  & \textbf{.} & \textbf{.} & \textbf{.} \\
\textbf{row 692}   &  {0}  & 21592.6878 &  302.5685 &  597.1364 \\
\textbf{row 693}   &  {1}  & {0} &  268.3588 &  329.3055 \\
\textbf{row 694}   &  {1}  & 79.8913 & 260.0223 & 453.6815 \\
\textbf{row 695}   &  {1}  & 275.8172 & 464.9036 & 457.9410 \\
\textbf{.}  &  \textbf{.}  & \textbf{.} & \textbf{.} & \textbf{.} \\
\textbf{.}  &  \textbf{.}  & \textbf{.} & \textbf{.} & \textbf{.} \\
\textbf{row 1320}   &  {1}   & {21600} & 474.0779 & 697.9201 \\
\textbf{.}  &  \textbf{.}  & \textbf{.} & \textbf{.} & \textbf{.} \\
\textbf{.}  &  \textbf{.}  & \textbf{.} & \textbf{.} & \textbf{.} \\
\textbf{row 70281}   &  {99}  & {0} &  327.1091 &  598.0677 \\
\textbf{row 70282}   &  {99}  & 141.9900 &  464.2400 &  603.2721 \\
\textbf{row 70283}   &  {99}  & 198.8276 & 472.7828 & 505.9701 \\
\textbf{.}  &  \textbf{.}  & \textbf{.} & \textbf{.} & \textbf{.} \\
\textbf{.}  &  \textbf{.}  & \textbf{.} & \textbf{.} & \textbf{.} \\
\textbf{row 70859}   &  {99}  & {21600} & 255.2838 & 574.6977 \\
\bottomrule
\end{tabular}
\vspace{-5mm}
\end{table}

\begin{figure}[h]
    \centering
    \vspace{-3mm}
    \includegraphics[width=0.75\linewidth]{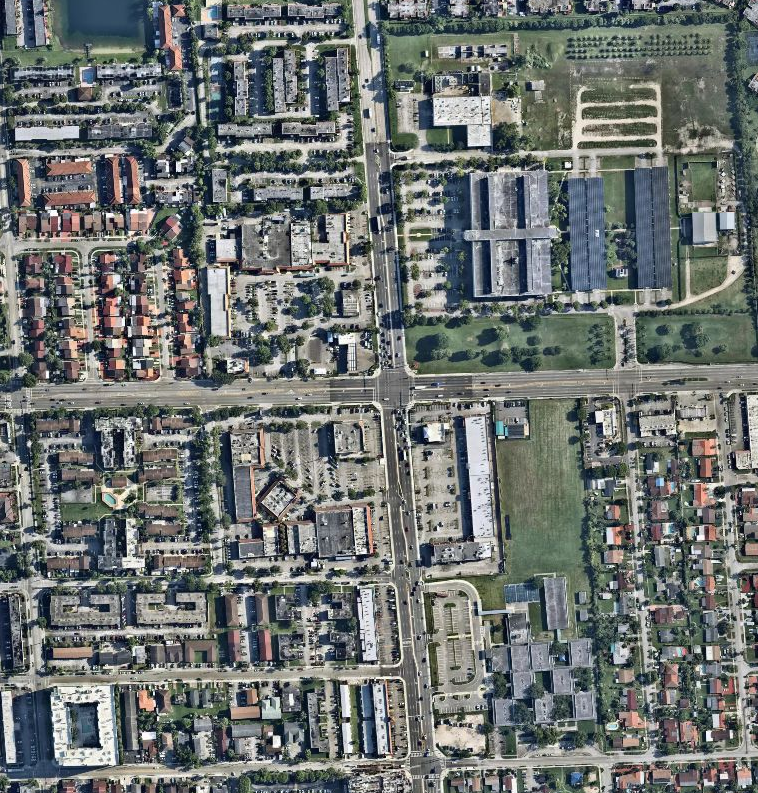}
    \vspace{-3mm}
    \caption{The map of FIU's Engineering Center and its surroundings}
    \vspace{-2mm}
    \label{fig:FIU_map}
\end{figure}

In short, a scenario includes nodes' coordinates at different times for the duration of the scenario. Thus, a scenario is essentially a network topology that changes over a period of time. We can also input real maps to the software to create more realistic scenarios. In this direction, we first chose a specific neighborhood around our school, Florida International University (FIU). Then, we extracted the map of FIU's Engineering Center Campus and its surroundings using the Java OpenStreetMap Tool (JOSM)\footnote{\url{https://josm.openstreetmap.de/}}. The map is shown in Fig. \ref{fig:FIU_map}. The bounding box of our map in longitude/latitude format is [\textit{-80.3733, 25.7647, -80.3653, 25.7722}]. With this map, we generated a number of mobility scenarios on the BonnMotion software using the Map-based Self-Similar Least-Action Walk (MSLAW) \cite{schwamborn2013introducing} algorithm. MSLAW was a suitable option for our case since it best captures moving people in a neighborhood and also it can take real maps as inputs. The scenario duration was 6 hours, the min and max speed of the pedestrians were 0.5 m/s and 2 m/s respectively. For the number of people (users), we chose 100, 200 and 300 where we created 40 scenarios for each for statistical significance.

Next, we preprocessed BonnMotion scenario files. The scenarios created by BonnMotion have different time values for each node as can be seen in Table \ref{tab:scenario}. In order to get the network topology at a specific time, we need all the nodes to have the identical time values. To achieve this, we grabbed the x-y coordinates of each node every 10 minutes from the scenario file and recorded the resulting data in a new \textit{.csv} file. In this way, every node had \textit{6 hours/10 minutes = 36} rows in the preprocessed scenario file.

With the preprocessed scenarios at hand, we could then create the mesh and LN topologies for the simulation. Each preprocessed scenario has 36 mesh topologies in it. To calculate a mesh topology from a scenario file for a given time, we first need to calculate the distances between all the nodes for all times. This is a costly operation to do at simulation run time, thus we opted for precomputing these distances and saving them into a \textit{.csv} file for each scenario. In this way, we can just read these files in the simulation run time and calculate the mesh topologies much faster. After this step, we also created the mobility-aware mesh topology for each scenario using Algorithm \ref{algo:algo1}. For the value of $d$, we used 90 meters to keep a conservative value for a potential IEEE 802.11n/ac coverage considering urban environments \cite{elkassabi2022experimental}. For $k$, we tried the following values: (\textit{2, 3, 4, 5, 6, 7, 8, 9, 10}). When $k$ was greater than 5, we started getting disconnected graphs for mobility-aware mesh topology. Values less than 5 on the other hand generated topologies with too many links which is not ideal. Thus, we chose 5 for the value of $k$. However, still we got 3 scenarios that did not include some nodes in the mobility-aware mesh topology thus, we performed the experiments with 37 scenarios instead.

\begin{figure*}[h!]
\centering
\begin{minipage}{.33\linewidth}
  \centering
  \includegraphics[width=\linewidth]{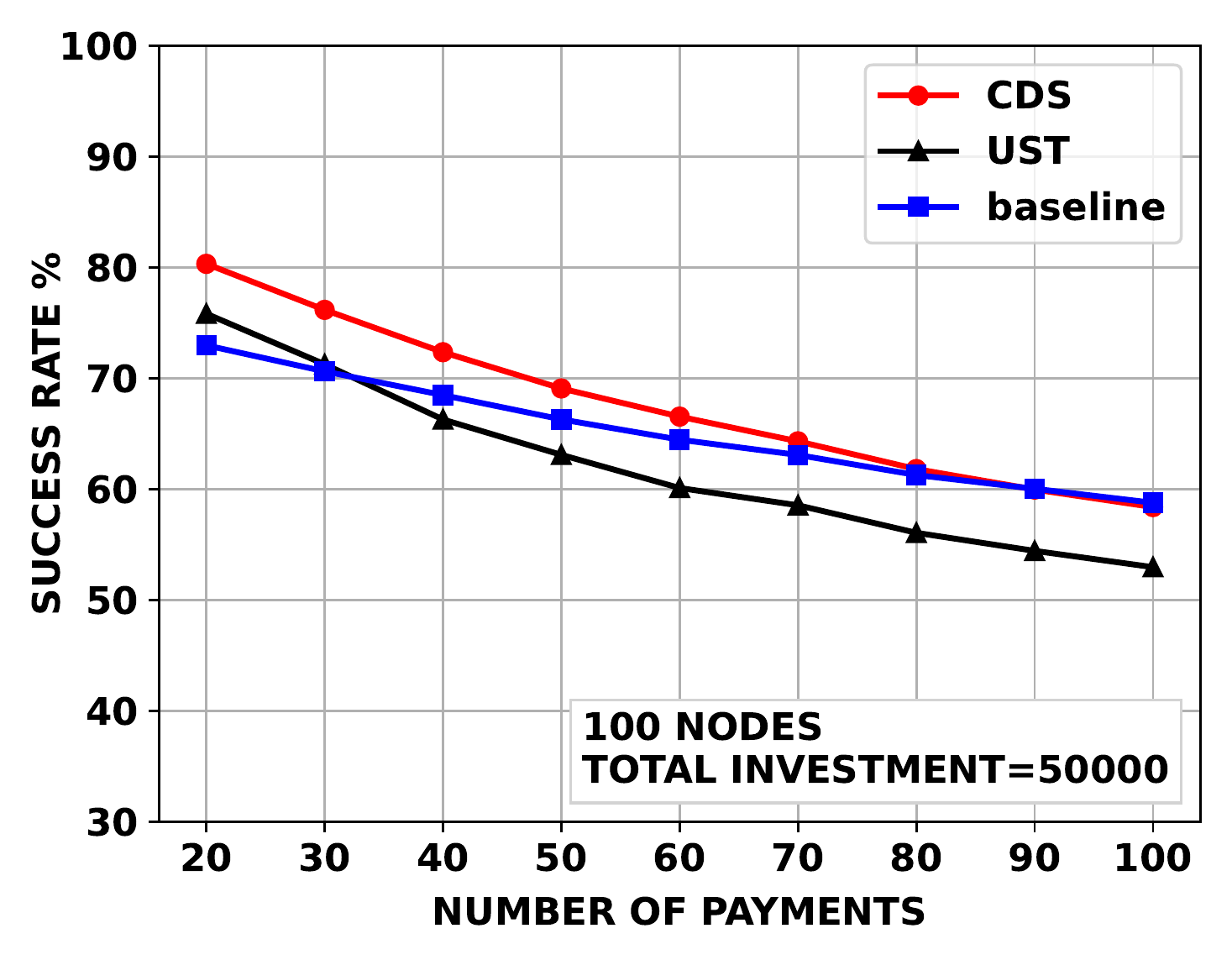}
  \vspace{-8mm}
  \caption{Success rate for 100 nodes.}
  \label{fig:result_a}
\end{minipage}%
\begin{minipage}{.33\linewidth}
  \centering
  \includegraphics[width=\linewidth]{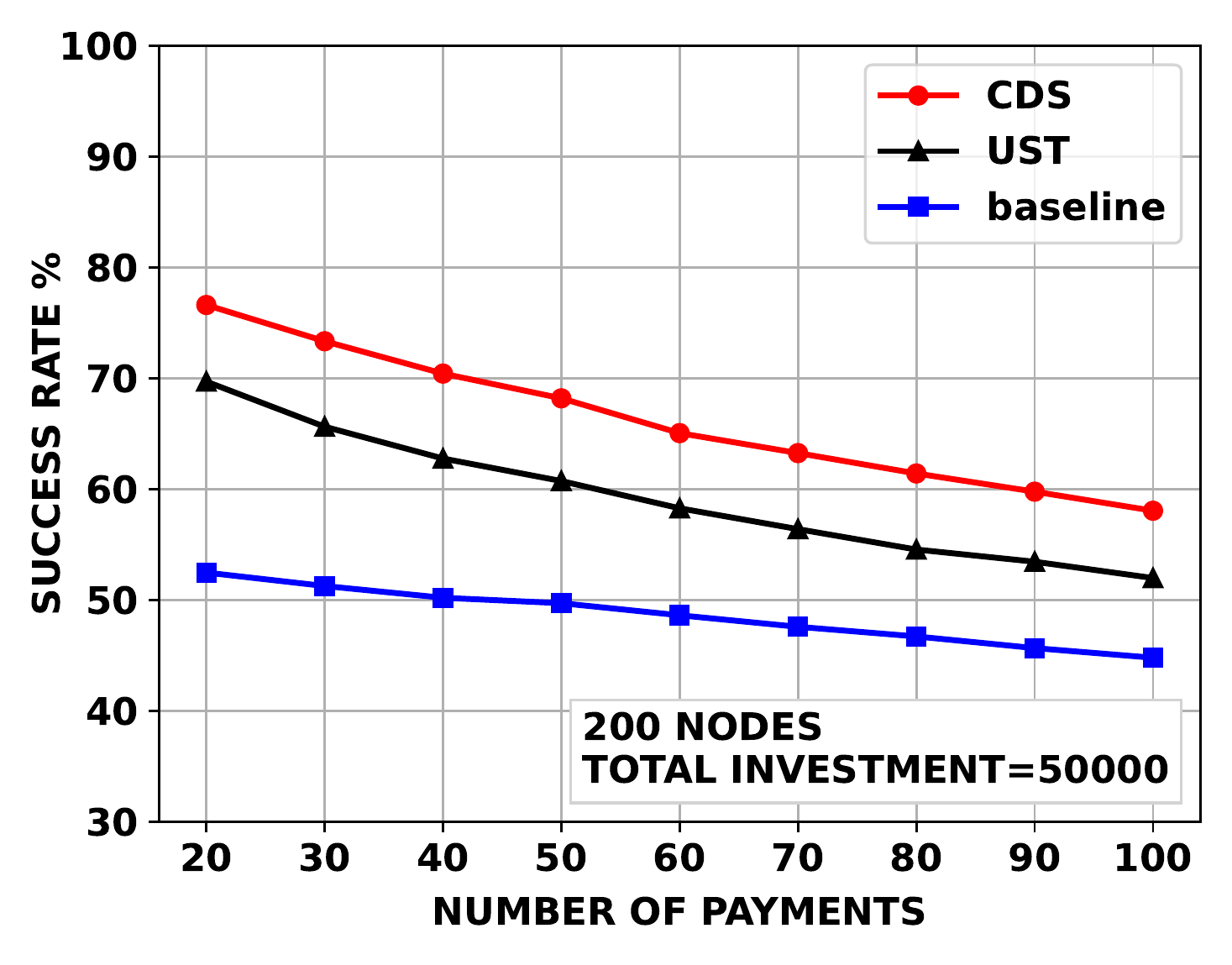}
  \vspace{-8mm}
  \caption{Success rate for 200 nodes.}
  \label{fig:result_b}
\end{minipage}%
\begin{minipage}{.33\linewidth}
  \centering
  \includegraphics[width=\linewidth]{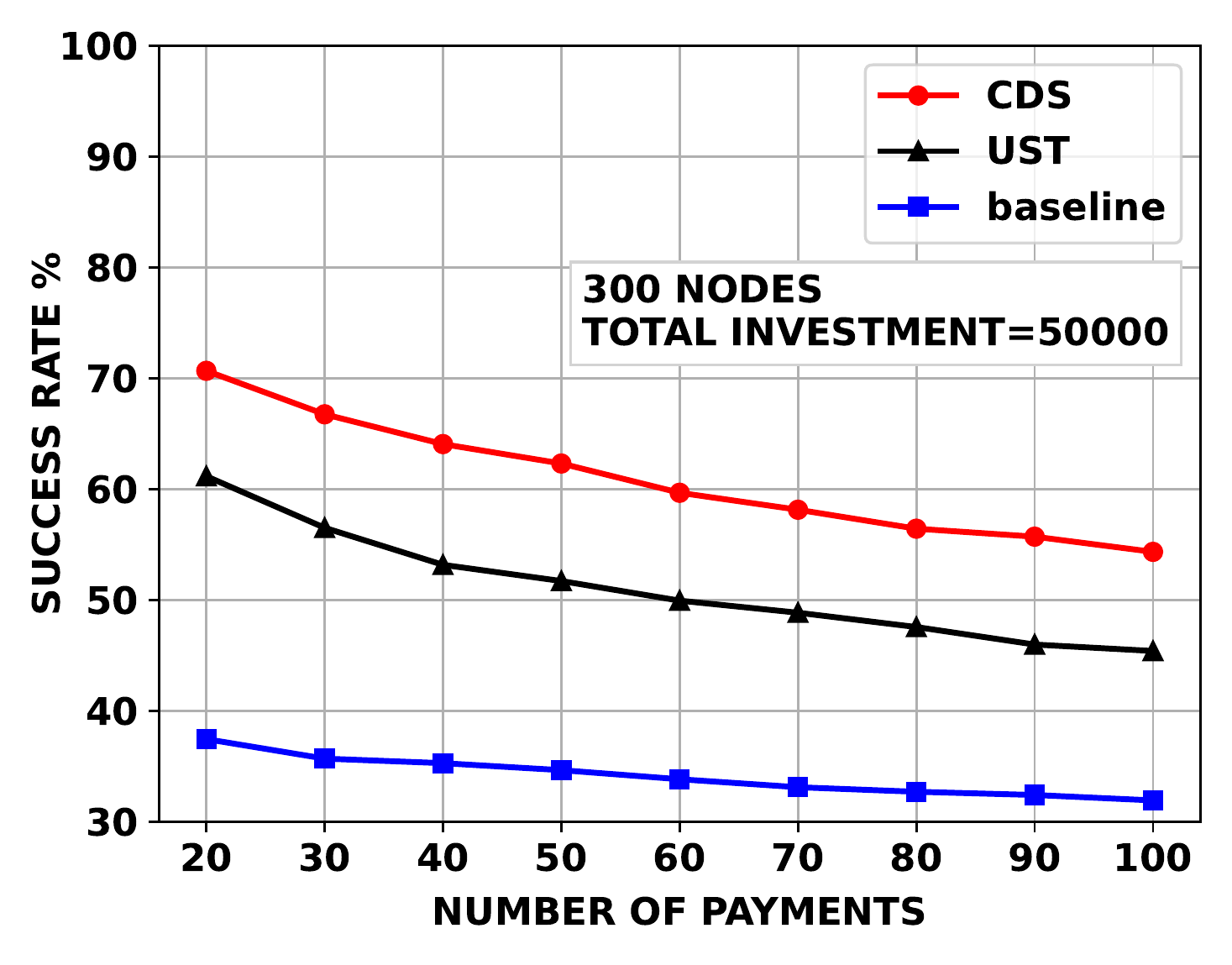}
  \vspace{-8mm}
  \caption{Success rate for 300 nodes.}
  \label{fig:result_c}
\end{minipage}
\end{figure*}

\begin{figure*}[h!]
\centering
\vspace{-5mm}
\begin{minipage}{.33\linewidth}
  \centering
  \includegraphics[width=\linewidth]{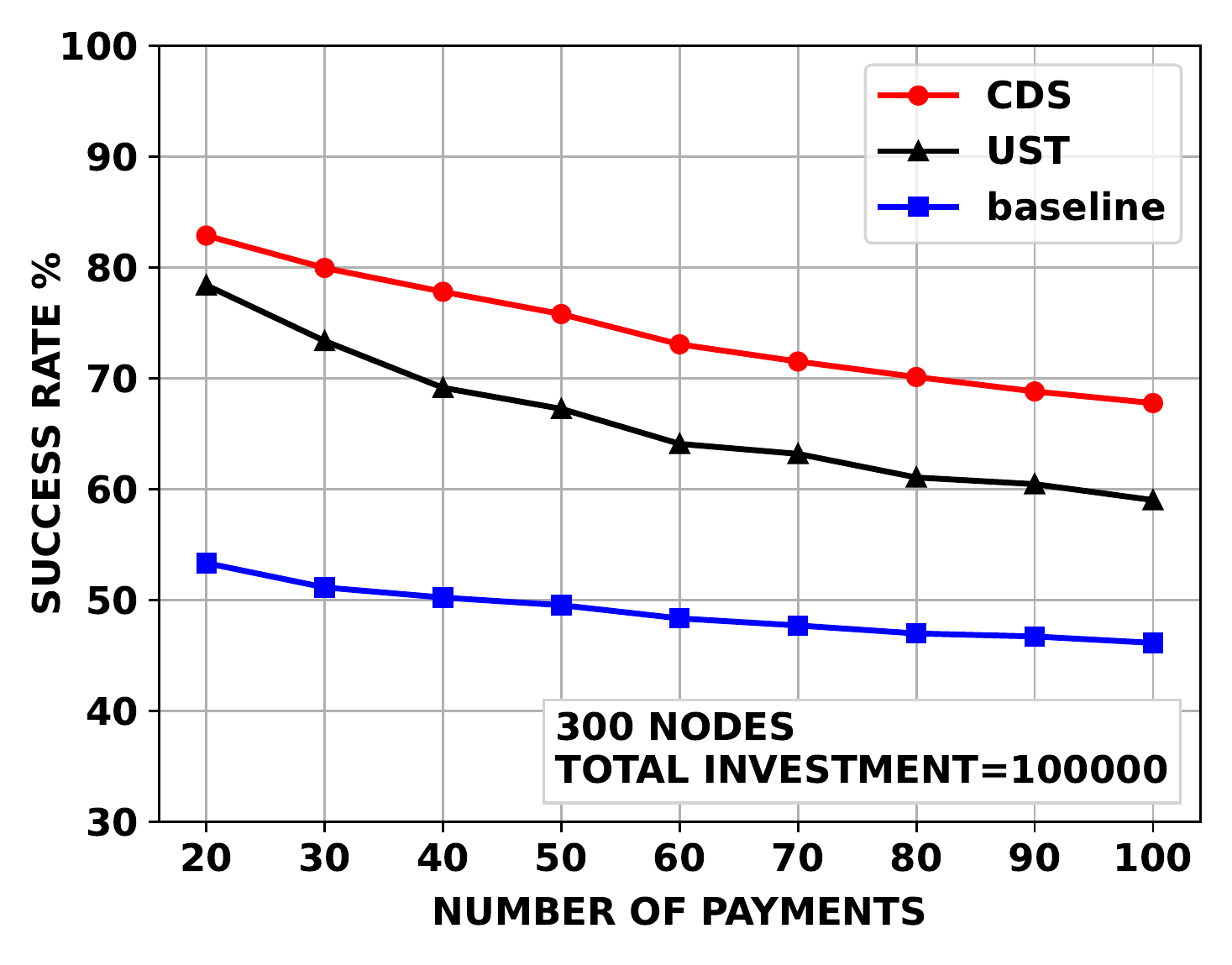}
  \vspace{-8mm}
  \caption{Success rate for 100000 satoshi.}
  \label{fig:result_d}
\end{minipage}%
\begin{minipage}{.33\linewidth}
  \centering
  \includegraphics[width=\linewidth]{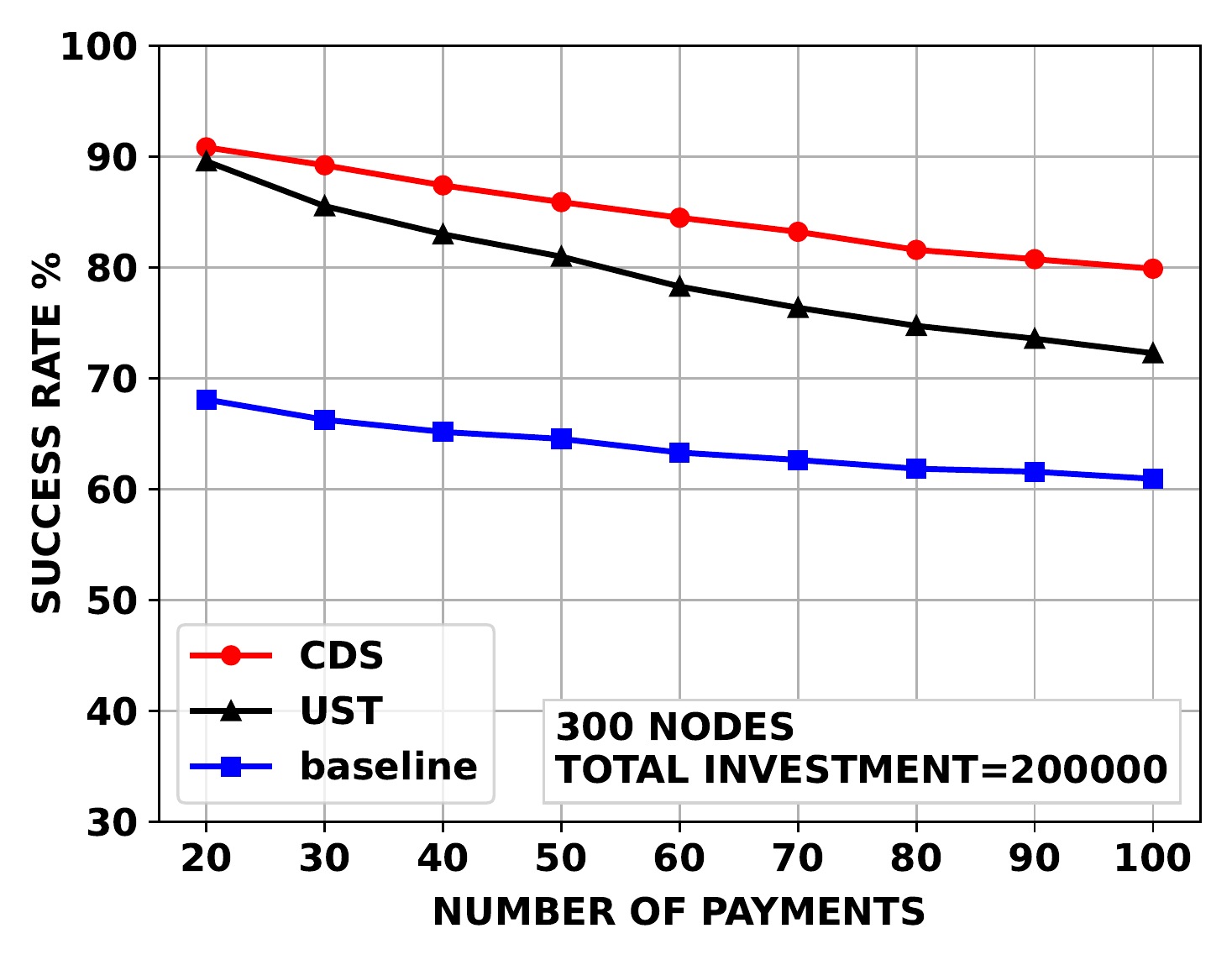}
  \vspace{-8mm}
  \caption{Success rate for 200000 satoshi.}
  \label{fig:result_e}
\end{minipage}%
\begin{minipage}{.33\linewidth}
  \centering
  \includegraphics[width=\linewidth]{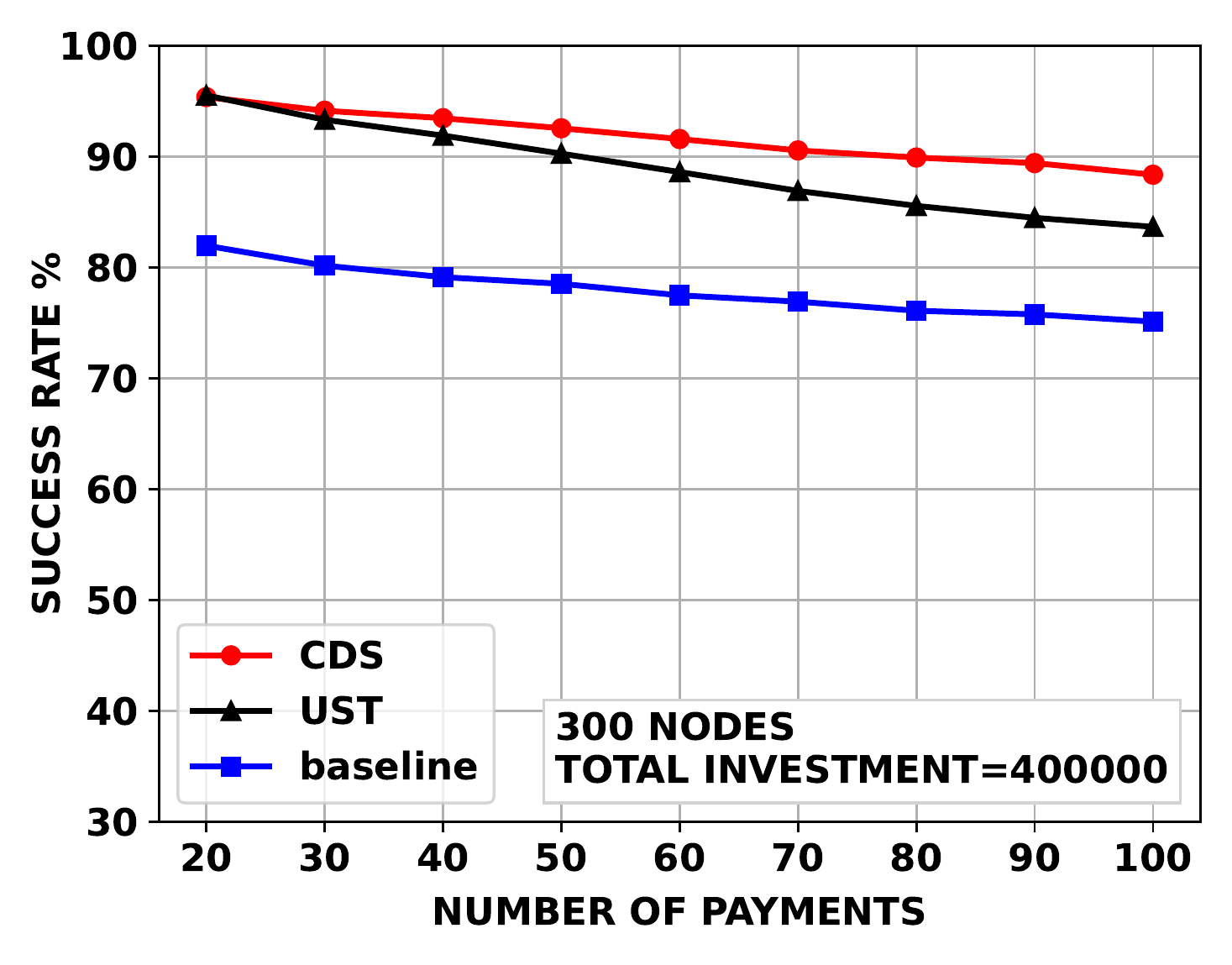}
  \vspace{-8mm}
  \caption{Success rate for 400000 satoshi.}
  \label{fig:result_f}
\end{minipage}
\vspace{-5mm}
\end{figure*}

\subsection{Simulation Setup, Metrics and Baselines}

\noindent \textbf{\textit{Simulation Setup}}: Our aim in the simulations is to perform \textit{N} arbitrary LN payments every 10 minutes between \textit{M} nodes. Amount \textit{A} of a payment is arbitrarily selected from the following list of values: (\textit{1, 5, 10, 20, 50, 100}) satoshi. Across the simulations, we varied \textit{N} with the following list of values: (\textit{20, 30, 40, 50, 60, 70, 80, 90, 100}). For the number of nodes \textit{M} in the simulations, we tested the following values: (\textit{100, 200, 300}). Basically, they match the number of nodes we initially input to the MSLAW algorithm during the scenario generation. As mentioned before, we used 37 scenarios for each value of \textit{M}. To have a fair comparison across the simulations, the random selection of $N$ payments with random amounts $A$ was done using fixed seed values. For channel capacities, we equally distributed a \textit{total investment amount} to all the channels in a simulation. The following total investment amounts were used in the simulations: (\textit{50000, 100000, 200000, 300000, 400000}) satoshi. For example, if there were 100 channels in a specific simulation and the total investment amount was 100,000 satoshi, then each channel was funded with 1,000 satoshi. Essentially, depending on the total investment and the total number of channels, channel capacities across the simulations vary. The reason for keeping the total investment amount fixed across simulations is to see the effect of other simulation parameters more clearly and fairly. For the MCDS algorithm, we used an existing Python implementation\footnote{\url{http://sparkandshine.net/en/calculate-connected-dominating-sets-cds/}} of the non-distributed MCDS algorithm proposed by Rai et al. \cite{rai2009new}. For generating USTs, we used \textit{DPPy} package in Python\footnote{\url{https://dppy.readthedocs.io/en/latest/exotic_dpps/ust.html}} which has an implementation of the Wilson's algorithm \cite{propp1998get} for general graphs.

\noindent \textbf{\textit{Metrics}}: We use the \textit{payment success rate} as our main metric on LN topologies which shows the percentage of transactions that were successfully received by the recipients. This metric depends on many parameters such as the total investment amount, number of payments every 10 minutes, number of nodes in the network and more importantly the selected channel assignment approach.

\noindent \textbf{\textit{Baseline}}: To test the effectiveness of the CDS and ST approaches, we need a baseline approach. The most intuitive candidate was the mobility-aware mesh topology which can be directly used as an LN topology. Thus, for the baseline approach, we opened an LN channel for each link in the mobility-aware mesh topology of the respective mobility scenario.

\subsection{Simulation Results}
We conducted several experiments whose results are shown in Fig. \ref{fig:result_a}-\ref{fig:result_f}. We discuss each scenario in details below:

\noindent \textbf{\textit{Impact of the number of nodes}}: Looking at Figures \ref{fig:result_a}, \ref{fig:result_b}, \ref{fig:result_c}, we can see that increasing the number of nodes in the network while keeping total investment fixed negatively affects the success rate. This is because, when there are more nodes in the network, there are also more channels. Since the total investment is kept the same, the channel capacities are lower on the network with more nodes. Thus, if we do not want to sacrifice on the success rate, we need to invest more money into the network when there are more users. This is actually not an issue in real life since each new user fund their channels with their own funds. Essentially, total investment automatically increases when a new user joins the network. Regardless of the number of nodes, increased number of payments deplete the channels quickly and thus there is a reduction in success rate for all approaches.

\noindent \textbf{\textit{Impact of total investment}}:  We can see from Figures \ref{fig:result_d}, \ref{fig:result_e}, \ref{fig:result_f} that increasing the total investment increases the success rates for all approaches dramatically. Specifically, when the total investment was increased from 50000 satoshi to 400000 for 300 nodes, the success rates reached up to \%95 for CDS, \%95 for UST, and \%82 for the baseline approach. Such increase in the success rates makes a lot of sense because by increasing the total investment, we are essentially putting more funds in each channel which helps the channels last longer and do not get depleted as quick. We see that even when the total investment is low (i.e., 50,000 satoshi), more than half of the payments in the network are successful for almost all cases.

\noindent \textbf{\textit{Impact of using CDS and UST}}: 
The most interesting outcome of the results are on the used approach. We can see from almost all the results that CDS consistently performs the best while UST comes after CDS and baseline approach performs the worst. To understand why CDS and UST have much higher success rate than the baseline, we need to look at the number of channels in each case which are shown in Table \ref{tab:num_channels}. These results are an average from 37 different mobility scenarios we used for the simulations.

As can be seen, baseline approach has too many channels compared to the CDS and UST. Since, we are keeping the total investment constant, the channel capacities in baseline experiments are very low compared to CDS and UST experiments. This in turn causes the payments to fail frequently due to insufficient capacity on the channels. The only exception was the case with Fig. \ref{fig:result_a} where UST performed worse than baseline even though its number of channels is lower. This can be attributed to the fact that there are much less users and alternative routes in this particular case. The channel capacities can be quickly depleted using same the routes for CDS and UST especially when the number of transactions increase. In case of baseline, there may be other routes that still allow successful transactions. When we increase the number of users to 300 in Fig. \ref{fig:result_d}, we see that the trends change completely since there are more users to offer more routes and thus baseline approach suffers from low channel capacities. Thus, it can be concluded that having fewer number of channels with higher capacities is more effective.

\begin{table}[h]
 \vspace{-3mm}
  \begin{center}
    \caption{Average Number of Channels for Different Approaches from 37 Different Scenarios}
     \vspace{-2mm}
    \label{tab:num_channels}
    \resizebox{0.85\linewidth}{!}{
    \begin{tabular}{|c|c|c|c|}
      \hline
      &   \multicolumn{3}{c|}{\textbf{Average Number of Channels}}    \\ 
     \hline 
     \textbf{Approach}  &  \textbf{100 nodes} & \textbf{200 nodes} & \textbf{300 nodes} \\
     \hline 
       CDS       &  99   &  199  &  299  \\ 
      \hline
       UST       &  99   &  199  &  299  \\
      \hline
       Baseline  &  665  &  2876 &  6381  \\ 
     \hline
     
    \end{tabular}
  }
  \vspace{-3mm}
  \end{center}
\end{table}

\begin{table*}[htb]
  \begin{center}
    \caption{Percentages of Payment Failures for a Total of 3,596,400 Payments from 45 Simulations}
    \vspace{-2mm}
    \label{tab:failures}
    \resizebox{0.72\linewidth}{!}{
    \begin{tabular}{|c|c|c|c|c|c|c|}
      \hline
      &   \multicolumn{3}{c|}{\textbf{Not Enough Capacity}}  & \multicolumn{3}{c|}{\textbf{No Mesh Path}}   \\ 
     \hline 
     \textbf{Approach}  &  \textbf{100 nodes}   & \textbf{200 nodes}   & \textbf{300 nodes}    & \textbf{100 nodes}   & \textbf{200 nodes}   & \textbf{300 nodes}  \\
      \hline
       CDS       &  \%15.45  &  \%16.27  &  \%19.2   &  \%2.71   &  \%2.43 &  \%2.18  \\ 
      \hline
       UST       &  \%19.14  &  \%20.83  &  \%25.38  &  \%2.86   &  \%2.63 &  \%2.47  \\
      \hline
       Baseline  &  \%15.17  &  \%26.77  &  \%40.20   &  \%2.26   &  \%1.75  &  \%1.35    \\ 
     \hline
    \end{tabular}
  }
  \vspace{-6mm}
  \end{center}
\end{table*}

In general, when we compare CDS and UST, we see that CDS has higher success rate than UST in all experiments. To understand why this is the case, we checked several properties of the CDS and UST graphs. First thing we noticed was CDS graphs have vertices with high degrees (i.e., vertices with many edges) while the vertex degrees in UST graphs were almost uniform. This causes payments to take longer paths in UST case which results in payments to fail more frequently due to channels with insufficient capacities along the payment path. CDS case on the other hand essentially has vertices acting as hubs which can relay the payments to recipients. As we predicted, this can be attributed to the \textit{closeness centrality} (CC) feature for CDS-based topologies which measures how short the shortest paths are from a node $i$ to all other nodes: $CC(i) = N-1 / \sum_j d(i,j) $ where $d(i,j)$ is the length of the shortest path between nodes $i$ and $j$ and $N$ is the number of nodes. Due to the way we generate CDS-based topologies, there are hub nodes which enable closeness centrality to be higher.

\noindent \textbf{\textit{Analysis of failed payments}}: 
To better understand the performance difference between CDS, UST and baseline approaches, we also analyzed the percentage for failed payments in Table \ref{tab:failures}. These results presented are cumulatively calculated from all of the simulations. Total number of payments for each case was 3,596,400.

Looking at the results, first thing we notice is that most payments fail because of not having enough capacity on the channels. The rest of the failures are due to not having paths on the mesh topology itself which are insignificant (i.e., only comprise \%2-3 of all payments and approaches to zero when we have more nodes). Note that we do not have failures due to not having paths on the LN topologies since we made sure all LN topologies we generated are connected.

Unarguably, the most important observation from these results is the effectiveness of CDS and UST approaches on reducing the payment failures from not having enough channel capacities. We see an improvement around up to \%50 over the baseline. Overall, these results show that with the right number of nodes and investment amount, we can achieve a success rate close to 100\% when using CDS-based approach.

\section{Conclusion}
\label{sec:conclusion}

In this paper, we proposed using mobile community wireless mesh networks to power offline LN payments so that people in a community/neighborhood without active Internet connections can continue transacting among themselves. We showed the feasibility of such a setup on a IP-over-BLE star network and a full WiFi mesh network using 8 Raspberry Pis. Our experiments implied that the way the channels are opened have a huge impact on the overall success rate of the payments in the network. Additionally, assuming that the mesh users can also move around complicates channel opening even more. Thus, we proposed two channel assignment approaches taking into account the mobility of the users. First approach is based on connected dominating set concept and the second one utilizes uniform spanning tree. To test these approaches in a large-scale, we implemented a simulator and extensively analyzed overall success rate of the payments on different settings. Our results showed that, our proposed approaches work well and can achieve success rates up to \%95 on large mobile wireless mesh networks.

\bibliographystyle{IEEEtran}
\bibliography{references}

\end{document}